\documentclass[12pt]{article} 
\usepackage{fixltx2e,fix-cm}
\usepackage{amssymb}
\usepackage{amsmath}
\usepackage{graphicx}
\usepackage{subfigure}
\usepackage{makeidx}
\usepackage{multicol}
\usepackage[authoryear]{natbib}
\usepackage{bbm}
\usepackage[margin=1in]{geometry}

\makeindex

\begin{document}

\title{Instrumental Variable Quantile Regression} 
\author{Victor Chernozhukov\thanks{MIT, email: vchern@mit.edu}\quad Christian Hansen\thanks{University of Chicago, email: chansen1@chicagobooth.edu}\quad Kaspar W\"uthrich\thanks{University of California San Diego, email: kwuthrich@ucsd.edu}}
\maketitle

\newtheorem{theorem}{Theorem}
\newtheorem{acknowledgement}[theorem]{Acknowledgement}
\newtheorem{algorithm}[theorem]{Algorithm}
\newtheorem{axiom}[theorem]{Axiom}
\newtheorem{case}[theorem]{Case}
\newtheorem{claim}[theorem]{Claim}
\newtheorem{conclusion}[theorem]{Conclusion}
\newtheorem{condition}[theorem]{Condition}
\newtheorem{conjecture}[theorem]{Conjecture}
\newtheorem{corollary}[theorem]{Corollary}
\newtheorem{criterion}[theorem]{Criterion}
\newtheorem{definition}[theorem]{Definition}
\newtheorem{example}[theorem]{Example}
\newtheorem{exercise}[theorem]{Exercise}
\newtheorem{lemma}[theorem]{Lemma}
\newtheorem{notation}[theorem]{Notation}
\newtheorem{problem}[theorem]{Problem}
\newtheorem{proposition}[theorem]{Proposition}
\newtheorem{remark}[theorem]{Remark}
\newtheorem{solution}[theorem]{Solution}
\newtheorem{summary}[theorem]{Summary}
\newtheorem{assumption}{Assumption}

\begin{abstract}
This chapter reviews the instrumental variable quantile regression model of \cite{iqr:ema}.  We discuss the key conditions used for identification of structural quantile effects within this model which include the availability of instruments and a restriction on the ranks of structural disturbances.  We outline several approaches to obtaining point estimates and performing statistical inference for model parameters.  Finally, we point to possible directions for future research. 
\end{abstract}



\section{Introduction}\label{Sec: Introduction}

Empirical analyses often focus on understanding the structural (causal) relationship between an outcome, $Y$, and variables of interest, $D$.  In many cases, interest is not just on how $D$ affects measures of the center of the distribution $Y$ but also on other features of the distribution.  For example, in understanding the effect of a government subsidized saving program, one might be more interested in the effect of the program on the lower tail of the savings distribution conditional on individual characteristics than on the effect of the program on the mean of the savings distribution.  Quantile regression, as introduced by \cite{koenker:1978}, offers one useful way to estimate such effects and to summarize the impact of changes in $D$ on the conditional distribution of $Y$.

Of course, variables of interest are often endogenous or self-selected in observational data.  For example, individuals choose whether to participate in government subsidized savings plans.  Similarly, in trying to understand the demand relationship between quantity and price, one must face that prices and quantities are jointly determined.  Endogeneity of covariates renders conventional quantile regression inconsistent for estimating the causal effects of variables on the quantiles of outcomes of interest.  Instrumental variables (IV) provide a powerful tool for learning about structural effects in the presence of endogenous right-hand-side variables, and we focus this review on a generalization of the classical linear instrumental variables model to accommodate estimating structural quantile treatment effects (QTE) in the presence of endogenous covariates. 

We specifically focus on the instrumental variable quantile regression model developed in \cite{iqr:ema}.  \cite{iqr:ema} provide conditions under which structural QTE are nonparametrically identified through the use of instrumental variables.  The key identifying assumption is a condition that restricts how structural errors, which we will refer to as rank variables, vary across different potential states of the endogenous variables.  The simplest, though strongest, version of this condition is rank invariance which requires that individual ranks are invariant to the potential states of the endogenous variable.  Rank invariance is implied by many classical structural models which posit a single source of unobserved heterogeneity, and the framework developed in \cite{iqr:ema} is indeed a natural generalization of the classical structural simultaneous equation model, corresponding to a structural simultaneous equation model with non-additive errors.  

There are alternative sets of modeling assumptions that one could employ to build a quantile model with endogeneity. \cite{abadie} offer an approach within the local average treatment effect framework of \cite{late}. This approach differs from the framework discussed in this review in a few key respects.  The \cite{abadie} framework does not restrict the behavior of rank variables across potential treatment states and thus allows for essentially unrestricted heterogeneity of effects at the cost of only identifying QTE for the subpopulation of compliers.  To achieve identification without restricting structural errors, \cite{abadie} restrict attention to a setting with a binary endogenous treatment variable and impose a monotonicity restriction on the relation between the instrument and treatment. When the endogenous variable of interest is continuous, triangular models as in \cite{imbens:newey} provide another alternative framework for identifying and estimating QTE. As in \cite{abadie}, the \cite{imbens:newey} framework does not restrict the evolution of ranks across treatment states. Instead, it relies on monotonicity of the selection mechanism in a scalar disturbance. We refer interested readers to \cite{melly:handbook} in this handbook, which discusses the \cite{abadie} approach in detail and contains further comparative discussion of the two modeling frameworks, and to \cite{wuthrich:ivqr}, which establishes a connection between the estimands of both models under the assumptions of the \cite{abadie} framework. Section \ref{SubSec: Other Approaches} discusses the approach by \cite{imbens:newey} and compares it to the framework discussed in this chapter. 

We devote the remainder of this review to providing an overview of the model of \cite{iqr:ema} along with outlining approaches to estimating parameters and performing inference within this model.

\section{Model Overview}\label{Sec: Model}

\subsection{The Instrumental Variable Quantile Regression Model} 
The instrumental variables quantile regression (IVQR) model is developed within the conventional potential outcome framework. 
Potential real-valued outcomes, which vary among observational units, are indexed against potential treatment states $d \in \mathcal{D}$ and denoted $Y_d$. The potential outcomes $\{ Y_{d}\}$ are latent because, given the observed treatment $D$, the observed outcome for each observational unit is only one component
$$ 
Y := Y_{D}
$$ 
of the potential outcomes vector $\{Y_d\}$.  Note that we use capital letters to denote random variables and lower case letters to denote the potential values the random variables may take throughout this review.  We also do not explicitly state technical measurability assumptions as these can be deduced from the context.

The objective of causal or structural analysis is to learn about features of the distributions of potential outcomes $Y_d$. Of primary interest to us are the $\tau^{\text{th}}$ quantiles of potential outcomes under various potential treatment states $d$, conditional on observed characteristics $X=x$, denoted as 
$$ 
Q_{Y_d}(\tau | x) = q(\tau,d,x).
$$ 
We note that, after conditioning on observed characteristics $X=x$, each potential outcome $Y_d$ can be related to its quantile function $q(\tau,d,x)$ as
\begin{align}\label{representation}
Y_d =  q(U_d,d,x), \text{ where } U_d \sim U(0,1)
\end{align}
is the structural error term and (\ref{representation}) follows from the Fisher-Skorohod representation of random variables.  

Given the conditional quantiles of the potential outcomes, we are then interested in QTE which are given by the difference in $\tau^{\text{th}}$ quantiles of two different conditional potential outcomes $Y_{d_1}$ and $Y_{d_0}$:
$$
q(\tau,d_1,x) - q(\tau,d_0,x).
$$
These QTE may then be used to summarize the impact of variables of interest $D$ on the quantiles of potential outcomes as suggested in \cite{doksum:1974} and \cite{lehmann:1975}.

It is important to note that the structural error $U_d$ in (\ref{representation}) is responsible for heterogeneity of potential outcomes among individuals with the same observed characteristics $x$. This error term determines the relative ranking of observationally equivalent individuals in the distribution of potential outcomes given the individuals' observed characteristics, and thus we refer to $U_d$ as the rank variable. Because $U_d$ drives differences between observationally equivalent individuals, one may think of $U_d$ as representing some unobserved characteristic, e.g. ability or ``proneness,'' where we adopt the term proneness from \cite{doksum:1974} who uses the term as in ``prone to learn fast" or ``prone to grow taller".  This interpretation of the structural error makes quantile analysis an interesting tool for describing and learning the structure of heterogeneous treatment effects while accounting for unobserved heterogeneity; see \cite{doksum:1974}, \cite{heckman:etal}, and \cite{koenker:book}.  For example, consider a returns-to-training model, where $Y_d$'s are potential earnings under different training levels $d$, and $q(\tau,d,x)$ is the conditional earnings function which describes how an individual with training $d$, characteristics $x$, and latent ``ability" $\tau$ is rewarded by the labor market. The earnings function may differ for different levels of $\tau$, implying heterogeneous effects of training on earnings of people that have different levels of ``ability". For example, it may be that the largest returns to training accrue to those in the upper tail of the conditional distribution, that is, to the ``high-ability" workers.

In observational data, the realized treatment $D$ is often selected in relation to potential outcomes, inducing endogeneity. This endogeneity makes the conventional quantile regression of $Y$ on $D$ and $X$, which relies upon the restriction
$$
P[ Y \leq Q_{Y}(\tau | D, X) | D, X] = \tau \ \text { a.s., }
$$
inappropriate for measuring the structural quantile function $q(\tau,d,x)$ and thus for learning about QTE.  Indeed, the conditional quantile function, $Q_{Y}(\tau | d, x)$, solving these equations will generally differ from the structural quantile function of latent potential outcomes, $q(\tau,d,x)$, under endogeneity.  The IVQR model presented below provides conditions under which we can identify and estimate the quantiles of the latent potential outcomes through the use of instruments $Z$ that affect $D$ but are independent of potential outcomes by making use of the nonlinear quantile-type conditional moment restrictions
$$
P[ Y \leq q(\tau,D,X) | X, Z] = \tau \ \text { a.s. }
$$

Formally, the IVQR model consists of five key conditions (some are representations).
\begin{assumption}[IVQR Model] Consider a common probability space $(\Omega, F, P)$ and the set of potential outcome variables $(Y_d, d \in \mathcal{D})$, endogenous variables $D$, exogenous covariates $X$, and instrumental variables $Z$.  The following conditions hold \textit{jointly} with probability one:
\begin{quote}
\begin{itemize}
\item[\textbf{A1 }] \textsc{Potential Outcomes.}  Conditional on $X$ and for each $d$, $Y_{d} = q(U_d,d,X)$, where $\tau \mapsto q(\tau,d,X)$ is non-decreasing on $[0,1]$ and left-continuous and $U_d \ {\sim} \ U(0,1)$.
\item[\textbf{A2 }] \textsc{Independence.  } Conditional on $X$ and for each $d$, $U_{d}$ is independent of instrumental variables $Z$.
\item[\textbf{A3 }] \textsc{Selection.} $D := \delta (Z, X, \nu)$ for some unknown function $\delta$ and random vector $\nu$.
\item[\textbf{A4 }] \textsc{Rank Similarity.} Conditional on $(X, Z, \nu)$,  $\{U_d\}$ are identically distributed.  
\item[\textbf{A5 }] \textsc{Observables.} The observed random vector consists of $Y := Y_D$, $D$, $X$ and $Z.$
\end{itemize}
\end{quote}
\end{assumption}

The following theorem summarizes the main econometric implications of the model.

\begin{theorem}[Main Implications of the IVQR Model]\label{Theorem: Main} Suppose conditions A1-A5 hold. (i) Then we have for $U:= U_D$, with probability one,
\begin{align} \label{main implication}
Y = q(U,D,X), \ \   U \sim U(0,1) |X,Z.
\end{align}
(ii) If (\ref{main implication}) holds and $\tau \mapsto q(\tau,d,X)$ is strictly increasing for each $d$, then for each $\tau \in (0,1)$, a.s
\begin{align}\label{e1}
P\left[Y \leq q(\tau,D,X)|X, Z \right]  = \tau.
\end{align}
(iii)  If (\ref{main implication}) holds, then for any closed subset $I$ of $[0,1]$, a.s.
\begin{align}\label{e1b}
P (U \in I) \leq  P\left[ Y \in  q(I,D,X)|X, Z \right],
\end{align}
where $q(I,d,x)$ is the image of $I$ under the mapping $\tau \mapsto q(\tau,d,x)$.
\end{theorem}

The first result states that the main consequence of A1-A5 is a simultaneous equation model (\ref{main implication}) with non-separable error $U$ that is independent of $Z,X$, and normalized so that $U \sim U(0,1)$.  The second result considers econometric implications when $\tau \mapsto q(\tau,D,X)$ is strictly increasing, which requires that $Y$ is non-atomic conditional on $X$ and $Z$.  In this case, we obtain the conditional moment restriction (\ref{e1}).  This implication follows from the first result and the fact that 
$$
\{ Y \leq q(\tau,D,X) \} \text{ is equivalent to }  \{ U \leq \tau\}
$$
when $q(\tau,D,X)$ is strictly increasing in $\tau$.  The final result deals with the case where $Y$ may have atoms conditional on $X$ and $Z$, e.g. when $Y$ is a count or discrete response variable.  The first two results were obtained in \cite{iqr:ema}, and the third result
is in the spirit of results given in \cite{ChesherRosenSmolinski}, \cite{ChesherDiscrete}, and \cite{ChesherSmolinski}. 


The model and the results of Theorem \ref{Theorem: Main} are useful for two reasons. First, Theorem \ref{Theorem: Main} serves as a means of identifying QTE in a reasonably general heterogeneous effects model. Second, by demonstrating that the IVQR model leads to the conditional moment restrictions \eqref{e1} and \eqref{e1b}, Theorem \ref{Theorem: Main} provides an economic and causal foundation for estimation based on these restrictions.  

Equations \eqref{e1} and \eqref{e1b} implicitly define the identification region for the structural quantile function $(\tau,d,x) \mapsto q(\tau,d,x)$. The identification region for the case of strictly increasing $\tau \mapsto q(\tau,d,x)$ can be stated as the set $\mathcal{M}$ of functions $(\tau,d,x) \mapsto m(\tau,d,x)$ that satisfy the following relations, for all $\tau \in (0,1]$
\begin{equation}\label{e2}
P[Y < m(\tau,D,X)|X, Z ] = \tau \text{ a.s. }
\end{equation}
This representation of the identification region $\mathcal{M}$ is implicit. Without imposing additional conditions, statistical inference about  $q \in \mathcal{M}$ from (\ref{e2}) can be performed using weak-identification robust inference as described in \cite{iqr:joe2}, \cite{JunWeakIdIVQR}, \cite{SantosPartialIDIVQR}, or \cite{fsqr}. Section \ref{Subsec: Identification} discusses conditions under which point identification is obtained; and we mainly focus on the point-identified case in discussing estimation and inference in this review.  

The identification region for the case of weakly increasing $\tau \mapsto q(\tau,d,x)$ can be stated as the set $\mathcal{M}$ of functions
$(\tau,d,x) \mapsto m(\tau,d,x)$ that satisfy the following relations:  For any closed subset $I$ of $(0,1]$,
\begin{align*}
P (U \in I) \leq  P\left[ Y \in m(I,D,X) |X, Z \right]  \text{ a.s.},
\end{align*}
where $m(I,D,X)$ is the image of $I$ under the mapping $\tau \mapsto m(\tau,D,X)$.   The inference problem here falls in the class of conditional moment inequalities and approaches such as those described in \cite{AndrewsShi} or \cite{CLRIntersectionBounds} can be used.  


\subsection{Conditions for Point Identification} \label{Subsec: Identification}

Here we briefly discuss the key conditions under which the moment equations (\ref{e1}) point identify the structural quantile function $q(\tau,d,x)$. We focus on the simplest case where $D\in \{0,1\}$ and $Z\in \{0,1\}$ and refer to \cite{iqr:annualreview} for more details and extensions to multivalued and continuous $D$ and to \cite{iqr:joe} for a discussion of identification in linear-in-parameters models. The following analysis is conditional on $X=x$, but we suppress this dependence for the ease of notation. 

It follows from Theorem \ref{Theorem: Main} that there is at least one function $q(\tau,d)$ that solves $P\left[Y\le q(\tau,D)|Z \right]=\tau$ a.s. The function $q(\tau,d)$ can be equivalently represented by a vector of its values $q=\left(q(\tau,0),q(\tau,1) \right)^\prime$. Therefore, for vectors of the form $y=\left(y_0,y_1 \right)^\prime$, we have a vector of moment equations
\begin{align*}
\Pi(y):= \left(P\left[Y\le y_D|Z=0 \right]-\tau, P\left[Y\le y_D|Z=1 \right]-\tau\right)^\prime,
\end{align*}
where $y_D:=(1-D)\cdot y_0+D\cdot y_1$. We say that $q(\tau,d)$ is identified in some parameter space, $\mathcal{L}$, if $y=q$ is the only solution to $\Pi(y)=0$ among all $y\in \mathcal{L}$. Define the Jacobian $\partial\Pi(y)$ of $\Pi(y)$ with respect to $y=\left(y_0,y_1\right)^\prime$ as
\begin{align} 
\partial\Pi(y):=&\bigg[\begin{matrix}f_{Y}\left(y_0|D=0,Z=0 \right)P\left[D=0|Z=0  \right] \\f_{Y}\left(y_0|D=0,Z=1 \right)P\left[D=0|Z=1  \right]  \end{matrix}\notag \\
&\begin{matrix}f_{Y}\left(y_1|D=1,Z=0 \right)P\left[D=1|Z=0  \right]\\ f_{Y}\left(y_1|D=1,Z=1 \right)P\left[D=1|Z=1\right]\end{matrix}\bigg]\notag \\
:=&\begin{bmatrix}f_{Y,D}\left(y_0,0|Z=0 \right) &f_{Y,D}\left(y_1,1|Z=0 \right)\\ f_{Y,D}\left(y_0,0|Z=1 \right) &f_{Y,D}\left(y_1,1|Z=1 \right)\end{bmatrix}
\end{align}
The key condition for point identification is full rank of $\partial\Pi(y)$ at $y=q$. This local identification condition can be extended to a global condition; see \cite{iqr:ema,iqr:annualreview}. 

Full rank of $\partial\Pi(y)$ requires the impact of $Z$ on the joint distribution of $(Y,D)$ to be rich enough. To illustrate, note that full rank of $\partial\Pi(y)$ is equivalent to $\det\left( \partial\Pi(y)\right)\ne 0$, which implies that 
\begin{align}
\frac{f_{Y,D}\left(y_1,1|Z=1 \right)}{f_{Y,D}\left(y_0,0|Z=1 \right)}>\frac{f_{Y,D}\left(y_1,1|Z=0 \right)}{f_{Y,D}\left(y_0,0|Z=0 \right)}\label{eq:monotone_likelihood}
\end{align}
(or the same condition with $>$ replaced by $<$). Inequality (\ref{eq:monotone_likelihood}) may be interpreted as a \emph{monotone likelihood ratio condition}. That is, the instrument $Z$ should have a monotonic impact on the likelihood ratio in (\ref{eq:monotone_likelihood}), which is generally stronger than the usual condition that $D$ is correlated with $Z$. Nevertheless, the full rank condition will be trivially satisfied in many useful contexts. For instance, if the instrument satisfies one-sided non-compliance (e.g., those not offered the treatment cannot receive that treatment), $P\left[D=1|Z=0 \right]=0$, so that the right-hand side of (\ref{eq:monotone_likelihood}) equals $0$, which makes (\ref{eq:monotone_likelihood}) hold trivially.


\subsection{Discussion of the IVQR Model} Condition A1 imposes monotonicity on the structural function of interest which makes its relation to $q(\tau,d,x)$ apparent.  Condition A2 states that potential outcomes are independent of $Z$, given $X$, which is a conventional independence restriction employed in nonlinear IV models.  Condition A3 provides a convenient representation of a treatment selection mechanism, stated for the purpose of discussion. In A3, the unobserved random vector $\nu$ is responsible for the difference in treatment choices $D$ across observationally identical individuals.  Dependence between $\nu$ and $\{U_d\}$ is the source of endogeneity that makes the conventional exogeneity assumption $U \sim U(0,1)|X,D$ break down.  This failure leads to inconsistency of exogenous quantile methods for estimating the structural quantile function.  Within the model outlined above, this breakdown is resolved through the use of instrumental variables.

The independence imposed in A2 and A3 is weaker than the assumption that both the disturbances $\{U_d\}$ in the outcome equation and the disturbances $\nu$ in the selection equation are \textit{jointly} independent of the instrument $Z$ which is maintained, for example, in \cite{abadie}. The assumption that structural errors $\{U_d\}$ and first-stage unobservables $\nu$ are jointly independent of instruments may be violated in practical examples.  For example, this condition would not hold when the instrument is measured with error as discussed in \cite{hausman:1977} or when the instrument is not assigned exogenously relative to the selection equation as in Example 2 in \cite{late}.

Condition A4 is the key restriction of the IVQR model.  This assumption restricts the variation in ranks across potential outcomes and is key for identifying the structural quantile function $q(\tau, d, x)$ and the associated QTE. The simplest, though  strongest, version of this condition is rank invariance which imposes that ranks $U_d$ do not vary with potential treatment states $d$: 
\begin{align} \label{invariance}
U_d =U \text{ for each } d \in \mathcal{D}.
\end{align}
Rank invariance is a strong condition that has been used in many interesting models without endogeneity such as \cite{doksum:1974}, \cite{heckman:etal}, and \cite{koenker:geling}.  Rank invariance implies that a common unobserved factor $U$, such as innate ability, determines
the ranking of a given person across treatment states.  For example, under rank invariance, people who are strong (highly ranked) earners without a training program ($d=0$) remain strong earners having done the training ($d=1$). Indeed, the earnings of a person with characteristics $x$ and
rank $U=\tau$ in the training state ``0" is $Y_0 = q(\tau, 0, x)$ and in the state ``1" is $Y_1 = q(\tau, 1, x)$; that is, the individual's rank, $\tau$, in the earnings distribution is exactly the same whether or not the person receives training.  Finally, note that Condition A3 is a pure representation under rank invariance as nothing restricts the unobserved component $\nu$ in this case.

While convenient, rank invariance seems too strong a condition for many applications as discussed, for example, in \cite{heckman:etal}.  Rank invariance maintains that an individual's rank in the outcome distribution under every possible state of the endogenous variables is exactly the same.  Thus, the potential outcomes $\{Y_d\}$ are jointly degenerate which allows identification of individual treatment effects even though no individual is ever observed in more than one state of the endogenous variable.  Rank invariance also rules out the possibility that there may be many unobserved factors that determine individual ranks which may be differentially relevant under different states of the endogenous variables.   

Rank similarity A4 relaxes these undesirable features of rank invariance by allowing the rank variables $\{U_d\}$ to change across $d$ in a way that reflects unobserved, asystematic variation in ranks across states of the endogenous variables while also providing sufficient structure to allow identification of QTE via the moment restrictions in Theorem \ref{Theorem: Main}.  More specifically, rank similarity A4 relaxes exact rank invariance by allowing ``slippages", in the terminology of \cite{heckman:etal},  in an individuals's rank away from some common level $U$.  Conditional on $U$, which may enter disturbance $\nu$ in the selection equation, and any other components of $\nu$ from the selection equation A3,  rank similarity yields that the slippages of ranks away from common level $U$ under different potential states of the endogenous variable, $U_d - U$, are identically distributed across $d \in \mathcal{D}.$  In this formulation, we implicitly assume that any selection of the state of the endogenous variables occurs without knowing the exact potential outcomes. That is, selection may depend on $U$ and even the distribution of slippages, but does not depend on the exact slippage $U_d-U$.  This assumption is consistent with many empirical situations where the exact latent outcomes are not known before receipt of treatment.  We also note that conditioning on appropriate covariates $X$ may be important to achieve rank similarity.  Finally, we note that rank similarity has testable implications. \cite{dongshen:ranksimilarity} and \cite{frandsen:ranksimilarity} exploit these conditions to develop tests of unconditional rank similarity, and their approaches could be extended to test some forms of conditional rank similarity.

\subsection{Examples}\label{Subsec: Examples}  We present two examples that highlight the nature of the model, its strengths, and its limitations.

\bigskip

\noindent{\textbf{Example 1}} (Demand with Non-Separable Error). The following is a generalization of the classic supply-demand example taken from \cite{iqr:joe}. Consider the model
\begin{align}\label{demand}
\begin{array}{llll}
 & Y_{p} = q\left(U, p\right), \\
 & \tilde Y_{p} = \rho\left(\mathcal{U}, p, z \right), \\
 & P \ \in \left\{p:  \rho\left(U, p, Z \right) =  q\left(\mathcal{U}, p \right) \right\},
\end{array}
\end{align}
where functions $q$ and $\rho$ are increasing in their first argument. The function $p \mapsto Y_p$ is the random demand function, and $p \mapsto \tilde Y_p$ is the random supply function. Additionally, functions $q$ and $\rho$ may depend on covariates $X$, but this dependence is suppressed.

Random variable $U$ is the level of demand and describes the demand curve at different states of the world. Demand is maximal when $U=1$ and minimal when $U=0$, holding $p$ fixed. Note that we imposed rank invariance (\ref{invariance}), as is typical in classic supply-demand models, by
making $U$ invariant to $p$.

Model (\ref{demand}) incorporates traditional additive error models for demand which have $Y_p = q(p) + \epsilon$ where $\epsilon = Q_{\mathcal{\epsilon}}(U)$. The model is much more general in that the price can affect the entire distribution of the demand curve, while in traditional models it only affects the location of the distribution of the demand curve.

The $\tau$-quantile of the demand curve $p \mapsto Y_p$ is given by $ p \mapsto  q(\tau, p). $ Thus, the curve $p \mapsto Y_p$ lies below the curve $p \mapsto  q(\tau, p)$  with probability $\tau$.  Therefore, the various quantiles of the potential outcomes play an important role in describing the distribution and heterogeneity of the stochastic demand curve. The QTE may be characterized  by $
\partial q(\tau, p)/ \partial p$ or by an elasticity $\partial \ln q(\tau, p)/ \partial \ln p.$ For example, consider the model $ q(\tau, p) = \exp\left( \beta(\tau) + \alpha(\tau) \ln p \right)$ which corresponds to a Cobb-Douglas model for demand with non-separable error $ Y_p = \exp ( \beta(U) + \alpha( U)\ln p ).$ The log transformation gives $ \ln Y_p = \beta(U) + \alpha( U) \ln p, $ and the QTE for the log-demand equation is given by the elasticity of the original $\tau$-demand curve $ \alpha(\tau) = \partial Q_{\scriptscriptstyle \ln Y_p}(\tau)/\partial \ln p = \partial \ln q(\tau, p) /\partial \ln p. $

The elasticity $\alpha(U)$ is random and depends on the state of the demand $U$ and may vary considerably with $U$. For example, this variation could arise when the number of buyers varies and aggregation induces a non-constant elasticity across the demand levels. \cite{iqr:joe2} estimate a simple demand model based on data from a New York fish market that was first collected and used by \cite{graddy}.  They find point estimates of the demand elasticity, $\alpha(\tau)$, that vary quite substantially from $-2$ for low quantiles to $-0.5$ for high quantiles of the demand curve.

The third condition in (\ref{demand}), $P \ \in \left\{p:  \rho\left(U, p, Z\right) = q\left(\mathcal{U}, p \right) \right\}$, is the equilibrium condition that generates endogeneity; the  selection of the clearing price $P$ by the market depends on the potential demand and supply outcomes. As a result,  we have a representation that is consistent with A3, $P = \delta ( Z, \nu), $ where $\nu$ consists of $U$ and $\mathcal{U}$ and may include``sunspot" variables if the equilibrium price is not unique. Thus what we observe can be written as
\begin{align}\label{simultaneous}
& Y  := q (U, P),  \  \  P := \delta ( Z, \nu), \ \ U
\text{ is independent of } Z.
\end{align}

Identification of the $\tau^{\text{th}}$ quantile of the demand function, $p \mapsto q(p, \tau)$ is obtained through the use of instrumental variables $Z$, like weather conditions or factor prices, that shift the supply curve and do not affect the level of the demand curve, $U$, so that independence assumption A2 is met.  Furthermore, the IVQR model allows arbitrary correlation between $Z$ and $\nu$. This property is important as it allows, for example, $Z$ to be measured with error or to be exogenous relative to the demand equation but endogeneous relative to the supply equation.

\bigskip

\noindent{\textbf{Example 2}} (Savings). \cite{401k} use the framework of the IVQR model to examine the effects of participating in a 401(k) plan on an individual's accumulated wealth.  Since wealth is continuous, wealth, $Y_d$, in the participation state $d \in \{0,1\}$ can be represented as
$$
Y_d = q(U_d, d, X),  \ \ U_d \sim U(0,1)
$$
where $\tau \mapsto q(\tau, d, X)$ is the conditional quantile function of $Y_d$ and $U_d$ is an unobserved random variable.  $U_d$ is an unobservable that drives differences in accumulated wealth conditional on $X$ under participation state $d$.  Thus, one might think of $U_d$ as the preference for saving and interpret the quantile index $\tau$ as indexing rank in the preference for saving distribution.  One could also model the individual as selecting the 401(k) participation state to maximize expected utility:
\begin{align}\label{roy}
\begin{split}
D & =  \arg \max_{d \in \mathcal{D}} \  E\left[ \ W \{ Y_d, d \} \Big | X,Z,\nu \right] \\
  & =  \arg \max_{d \in  \mathcal{D}} \ E\left[ \ W \{q (U_d, d, x), d\}  \Big | X,Z,\nu \right],
\end{split}
\end{align}
where $W\{Y_d,d\}$ is the random indirect utility derived under participation state $d$.  Of course, utility may depend on both observables in $X$ as well as realized and unrealized unobservables. Only dependence on $Y_d$ and $d$ is highlighted. As a result, the participation decision is represented by 
$$ 
D = \delta (Z, X, \nu), 
$$
where $Z$ and $X$ are observed, $\nu$ is an unobserved information component that may be related to ranks $U_d$ and includes other unobserved
variables that affect the participation state, and function $\delta$ is unknown. This model fits into the IVQR model with the independence condition A2 requiring that $U_d$ is independent of $Z$, conditional on $X$.

Under rank invariance \eqref{invariance} the preference for saving vector ${U_d}$ may be collapsed to a single random variable $U = U_0 = U_1.$ In this case, a single preference for saving is responsible for an individual's ranking across both treatment states. The more general rank similarity condition A4 relaxes the exact invariance of ranks $U_d$ across $d$ by allowing noisy, asystematic variations of $U_d$ across $d$, conditional on $(\nu,X,Z)$.  This relaxation allows for variation in rank across the treatment states, requiring only an ``expectational rank invariance."  Similarity implies that given the information in $(\nu,X,Z)$ employed to make the selection of treatment $D$, the expectation of any function of rank $U_d$ does not vary across the treatment states. That is, ex-ante, conditional on $(\nu,X,Z)$, the ranks may be considered to be the same across potential treatments, but the realized, ex-post, rank may be different across treatment states.

From an econometric perspective, the similarity assumption is nothing but a restriction on the unobserved heterogeneity component which precludes systematic variation of $U_d$ across the treatment states. To be more concrete, consider the following simple example where 
$$ 
U_d = F_{\nu+\eta_d}(\nu + \eta_d),
$$
where $F_{\nu+\eta_d}(\cdot)$ is the distribution function of $\nu+\eta_d$ and $\{\eta_d\}$ are mutually i.i.d. conditional on $\nu$, $X$, and $Z$. The
variable $\nu$ represents an individual's ``mean" saving preference, while $\eta_d$ is a noisy adjustment.  Clearly similarity holds in this case, $U_d \overset{d}= U_{d'}$ given $\nu$, $X$, and $Z$.  This more general assumption leaves the individual optimization problem (\ref{roy}) unaffected, while allowing variation in an individual's rank across different potential outcomes.

While we feel that rank similarity may be a reasonable assumption in many contexts, imposing rank similarity is not innocuous.  In the context of 401(k)
participation, matching practices of employers could jeopardize the validity of the similarity assumption.    To be more concrete, let $U_d = F_{\nu+\eta_d}(\nu + \eta_d)$ as before but let $\eta_d = d M$ for random variable $M$ that depends on the match rate and is independent of $\nu$, $X$, and $Z$.  Then conditional on $\nu = v$, $X$, and $Z$, $U_0 = F_\nu(v)$ is degenerate but $U_1 = F_{\nu+M}(v+M)$ is not.  Therefore, $U_1$ is not equal to $U_0$ in distribution.  Similarity may still hold in the presence of the employer match if the rank, $U_d$, in the asset distribution is insensitive to the match rate.  The rank may be insensitive if, for example, individuals follow simple rules of thumb such as target saving when they make their savings decisions.  Also, if the variation of match rates is small relative to the variation of individual heterogeneity
or if the covariates capture most of the variation in match rates, then similarity may be satisfied approximately.

\subsection{Comparison to Other Approaches} \label{SubSec: Other Approaches}

There are, of course, other assumptions that one could employ to build a quantile model with endogeneity. In this section, we briefly compare the IVQR framework to triangular models as in \cite{imbens:newey}; see \cite{chesher}, \cite{koenker:ma}, \cite{SLeeTriangularIVQR} and \cite{chernozhukov:cqiv} for related models and results. We also note that triangular models are related to the Rosenblatt transform; see for example the chapter by \citet{hallin:siman} in this handbook. A comparison between the IVQR model and the popular \cite{abadie} approach is provided in \citet{melly:handbook} in this handbook. 

The triangular model takes the form of a triangular system of equations
\begin{align*}
Y &= g(D,\epsilon), \\
D &= h(Z,\eta),
\end{align*}
where $Y$ is the outcome, $D$ is a continuous scalar endogenous variable, $\epsilon$ is a vector of disturbances, $Z$ is a vector of instruments with a continuous component, $\eta$ is a scalar reduced form error, and we ignore other covariates $X$ for simplicity.  It is important to note that the triangular system generally rules out simultaneous equations which typically have that the reduced form relating $D$ to $Z$ depends on a vector of disturbances. For example, in a supply and demand system, the reduced form for both price and quantity will generally depend on the unobservables from both the supply equation and the demand equation; see Example 1 in Section \ref{Subsec: Examples}.

Outside of $\eta$ being a scalar, the key conditions that allow identification of quantile effects in the triangular system are (a) the function $\eta \mapsto h(Z,\eta)$ is strictly increasing in $\eta$ and (b) $D$ and $\epsilon$ are independent conditional on $V$ for some observable or estimable $V$. The variable $V$ is thus the ``control function'' conditional on which changes in $D$ may be taken as causal. \cite{imbens:newey} use $V = F_{D|Z}(d,z) = F_{\eta}(\eta)$ as a control variable and show that this variable satisfies condition (b) under the additional condition that $(\epsilon,\eta)$ is independent of $Z$. Identification then proceeds as follows. Under the assumed monotonicity of $h(Z,\eta)$ in $\eta$, $D = h(Z,\eta)$ can be used to identify $V$. Using $V$ obtained in this first step, one may then construct the distribution of $Y|D,V$. Integrating over the distribution of $V$ and using iterated expectations, one has
\begin{align*}
\int F_{Y|D,V}(y|d,v)F_V(dv) &= \int 1(g(d,\epsilon) \le y)F_{\epsilon}(d\epsilon) \\
&= \textnormal{Pr}(g(d,\epsilon) \le y) := G(y,d)
\end{align*}
and the structural quantile function $Y_d$ can be obtained as $G^{-1}(\tau,d)$.

It should be emphasized that the triangular model is neither more nor less general than the IVQR model reviewed here. The key difference between the approaches is that the IVQR model uses an essentially unrestricted selection equation ($\nu$ may be vector valued) but requires monotonicity and a scalar disturbance ($U$) in the structural equation.  The triangular system on the other hand relies on monotonicity of the selection mechanism in a scalar disturbance ($\eta$) but does not restrict the unobserved heterogeneity in the outcome equation ($\epsilon$ may be a vector of disturbances). In addition, the triangular system, as developed in \cite{imbens:newey}, requires a more stringent independence condition in that the instruments $Z$ needs to be independent of both the structural disturbances, $\epsilon$, and the reduced form disturbance, $\eta$. That the approaches impose structure on different parts of the model makes them complementary with a researcher's choice between the two being dictated by whether it is more natural to impose restrictions on the structural function or the reduced form in a given application. 

Finally, we note that the triangular model and the IVQR model can be made compatible by imposing the conditions from the triangular model on the selection equation and the conditions from the IVQR model on the structural model.  \cite{torgovitsky:id} studies identification when both sets of conditions are imposed and shows that the requirements on the instruments may be substantially relaxed relative to the IVQR model or \cite{imbens:newey} in this case.

\section{Basic Estimation and Inference Approaches}\label{Sec: Estimation}

In this section, we present various approaches to estimating and doing inference for the parameters of the IVQR model under the leading case where $\tau \mapsto q(\tau,d,X)$ is strictly increasing.  We focus on linear-in-parameters structural quantile models at a single quantile of interest $\tau$:
\begin{align}\label{qsf: linear}
q(\tau,d,x) = d'\alpha_0(\tau) + x'\beta_0(\tau).
\end{align}
In (\ref{qsf: linear}), $\alpha_0(\tau)$ captures the causal effect of the endogenous variables $D$ on the $\tau^{\text{th}}$ quantile of the conditional distribution of potential outcomes $Y_d$ given $X = x$.  Similarly, $\beta_0(\tau)$ provides the causal effect of controls $X$ on the $\tau^{\text{th}}$ quantile of the conditional potential outcome distributions. We note that $D$ may also contain interactions of endogenous variables and covariates. Because $\alpha_0(\tau)$ is the chief object of interest in many studies, we focus most of our discussion on estimating and doing inference for $\alpha_0(\tau)$ treating $\beta_0(\tau)$ as a nuisance parameter. Note that in what follows we will often suppress the dependence of $\alpha_0(\tau)$ and $\beta_0(\tau)$ on the quantile level $\tau$.

In interpreting the parameters in (\ref{qsf: linear}), it is important to note that the quantile index, $\tau$, refers to the quantile of potential outcome $Y_d$ given that exogenous variables are set to $X = x$ and not to the unconditional quantile of $Y_d$.  For example, suppose that one of the control variables in the savings example in Section \ref{Subsec: Examples} is income.  An individual at the 10$^{\textnormal{th}}$ percentile of the distribution of $Y_d$ given an income of \$200,000, which is far above the median income, may not necessarily be at the low tail of the unconditional distribution of $Y_d$ as even a relatively low saver with a high level of income may still save substantially more than the median saver in the overall population, i.e., without conditioning on income; see \cite{FroelichMellyUnconditionalIVQR} for a further discussion of this point. In some applications, features of the conditional distribution are not the chief objects of interest and researchers are interested in effects of treatments on unconditional quantiles. Unconditional QTE can be obtained from the conditional quantile functions in three steps. First, obtain the conditional potential outcome distribution functions, $F_{Y_d}\left(y|x\right)$, as
$$
F_{Y_d}\left(y|x\right)=\int_0^1 \mathbbm{1}\left(d'\alpha_0(\tau) + x'\beta_0(\tau)\le y\right)d\tau,
$$
where $\mathbbm{1}(\cdot)$ is the indicator function that returns one when the expression inside the parentheses is true and zero otherwise. Second, the unconditional potential outcome distributions, $F_{Y_d}\left(y\right)$, are obtained by integrating $F_{Y_d}\left(y|x\right)$ with respect to the marginal distribution of covariates, $F_X(x)$:
$$
F_{Y_d}\left(y\right)=\int F_{Y_d}\left(y|x\right)dF_X(x).
$$
Finally, the unconditional $\tau$-QTE is given by $F^{-1}_{Y_{d_1}}\left(\tau\right)-F^{-1}_{Y_{d_0}}\left(\tau\right)$. This discussion suggests that given estimators of the parameters $\alpha_0(\tau)$ and $\beta_0(\tau)$ and the distribution of covariates $F_X(x)$, unconditional QTE can be estimated based on the plug-in principle; see for instance \cite{machado:mata}, \cite{melly:decomposition} or \cite{chernozhukov:counterfactual}.

Model (\ref{qsf: linear}) provides a simple and widely used baseline for discussion of estimation and inference.  Extending the discussion to allow for nonlinear parametric specifications of the potential outcome quantile functions or to estimation at a small number of quantile indices that are widely spaced is straightforward.  In some applications, we may be interested in understanding QTE across a range of quantile indices, say $\tau \in [\delta,1-\delta]$ for some $\delta > 0$.  \cite{iqr:joe} explicitly consider this case and provide uniform convergence results which allow for inference about a variety of hypotheses surrounding the behavior of QTE viewed as a function of $\tau$ such as tests of monotonicity of treatment effects or tests that treatment effects are uniformly 0 across a range of $\tau$.  Finally, we note that \cite{CIN:JoE}, \cite{HorowitzLeeNonparametricIVQR}, \cite{ChenPouzo2009}, \cite{ChenPouzo2012}, and \cite{GagliardiniScaillet} consider fully nonparametric approaches to estimating structural quantile models.

\subsection{Generalized Methods of Moments and Related Approaches}\label{Subsec: GMM}

The most direct way to estimate the parameters of the linear IVQR model is to note that the main implication of the model, equation (\ref{e1}), implies unconditional moment conditions
\begin{align}\label{eq: moment}
\text{E}\left[\left(\tau - \mathbbm{1}\left(Y - D'\alpha_0 - X'\beta_0 \leq 0\right)\right)\Psi \right] = 0
\end{align}
where $\Psi := \Psi(X,Z)$ is a vector of functions of the instruments and endogenous variables.\footnote{A natural choice of instruments would be $\Psi = (Z',X')'$ though the instruments and GMM weighting matrix could be chosen to produce a pointwise efficient procedure following \cite{chamberlain:iv}.}  Supposing that $\alpha_0$ is an $s \times 1$ vector and $\beta_0$ is a $k \times 1$ vector, a minimal necessary condition for identifying the model parameters will be $\dim\left(\Psi\right) = r \geq k + s$.

Let, for $\theta:= (\alpha, \beta)$ and $V:= (Y,D,X,Z)$,
$$
g_\tau (V, \theta) = \left(\tau - \mathbbm{1}\left(Y - D'\alpha - X'\beta \leq 0\right)\right)\Psi.
$$
With a given set of instruments, $\Psi$, and observables $\{V_i\}_{i=1}^{N} = \{Y_i,D_i,X_i,Z_i\}_{i=1}^{N}$, one may then form the sample analog of the right-hand-side of the equation (\ref{eq: moment}),
\begin{align}\label{eq: sample moment}
\widehat{g}_N(\theta) = \frac{1}{N}\sum_{i = 1}^{N} g_\tau(V_i, \theta), 
\end{align}
and estimate $\theta_0 = (\alpha_0',\beta_0')'$ by generalized method of moments (GMM) as
\begin{align}\label{eq: GMM obj}
\widehat\theta = (\widehat\alpha',\widehat \beta')' = \arg\min_{\theta \in \Theta} m_N(\theta) 
\end{align}
for 
$$
m_N(\theta) := N \widehat{g}_N(\theta)'\Omega_N \widehat{g}_N(\theta)
$$
where $\Omega_N$ is the GMM weighting matrix that will typically be set as 
$$
\Omega_N = \left ( \tau (1- \tau) \frac{1}{N} \sum_{i=1}^N \Psi_i \Psi_i'  \right)^{-1}.
$$ 
Maintaining sufficient conditions for point identification as in \cite{iqr:ema,iqr:joe,iqr:annualreview} and assuming that a suitable solution to the GMM optimization problem (\ref{eq: GMM obj}) can be found, asymptotic properties of $\widehat\theta(\tau)$ would then follow from standard results for GMM with non-smooth moment conditions as in \cite{newey:mcfadden}; see \cite{abadie:1995} and
\cite{chern:mcmc}. We note that if the GMM problem (\ref{eq: moment}) is overidentified, overidentification-type tests can be used to assess the joint validity of the underlying assumptions.

The chief difficulty in implementing estimation based on (\ref{eq: GMM obj}) is that the function being minimized is both non-smooth and non-convex in general.  We also note that in many applications, $s$ will be small, often one, but $k$ may be quite large.  Solving (\ref{eq: GMM obj}) then involves optimizing a non-smooth, non-convex function over $s + k$ arguments where $s + k$ may be quite large.  Directly solving this problem thus poses a substantial computational challenge and has led to the adoption of different approaches to estimating the parameters of the IVQR model.

Within the conventional GMM framework, one option is to take the quasi-Bayesian approach of \cite{chern:mcmc}; see also \cite{wang:yang} in this handbook for a review of subsequent work on related methods. The \cite{chern:mcmc} approach uses the GMM criterion function to form a ``quasi-likelihood'',
$$
L_N(\theta) = \exp\left ( - \frac{1}{2} N \widehat{g}_N(\theta)'\Omega_N \widehat{g}_N(\theta) \right),
$$
which when coupled with a prior density $\pi(\theta)$ over model parameters  $\theta$, defines a ``quasi-posterior'' density for $\theta$:
$$
\pi_N(\theta) = L_N(\theta)\pi( \theta)/ \int  L_N(\tilde \theta) d \pi(\tilde \theta) \propto L_N(\theta)\pi( \theta). 
$$

 Rather than try to solve the optimization problem (\ref{eq: GMM obj}), one can then use MCMC sampling to attempt to explore the implied quasi-posterior distribution.  \cite{chern:mcmc} show that measures of central tendency from the quasi-posterior, such as the quasi-posterior mean, 
 $$
\widehat\theta = (\widehat\alpha',\widehat \beta')' =  \int \theta d \pi_N(\theta)
 $$
 and quasi-posterior median  are consistent for model parameters with the same asymptotic distribution as the solution to (\ref{eq: GMM obj}).  \cite{chern:mcmc} also demonstrate that valid frequentist confidence intervals may be obtained by taking quasi-posterior quantiles. For example, a frequentist 95\% confidence interval may be constructed as by taking the 2.5 and 97.5 quantiles of the quasi-posterior distribution. This approach bypasses the need to optimize a non-convex and non-smooth criterion at the cost of needing to design a sampler that adequately explores the quasi-posterior in a reasonable amount of computation time.  
 
A second option is to directly smooth the GMM-criterion function as in \cite{kaplan2016}, building upon ideas in \cite{amemiya:1982} and \cite{horowitz:1998}. Specifically, one modifies the moment condition (\ref{eq: sample moment}) to
\begin{align}\label{eq: smooth moment}
\widehat{g}^{h_N}_N(\theta) = \frac{1}{N}\sum_{i = 1}^{N} \left(\tau - G_{h_N}\left(Y_i - D_i'\alpha - X_i'\beta \right)\right)\Psi_i,
\end{align}
by smoothing the indicator function, where $G_h(\cdot)$ denotes a smoothing function with smoothing parameter $h$.  $G_h(\cdot)$ can be defined as the survival function associated with any kernel function $K_h(\cdot)$, i.e. $G_h(u) = \int_{u}^{\infty} K_h(v)dv$, that satisfies regularity conditions provided in \cite{kaplan2016}.  One can then proceed to estimate model parameters by replacing $\widehat{g}_N(\theta)$ in (\ref{eq: GMM obj}) with $\widehat{g}^{h_N}_N(\theta)$ and applying any optimizer which is appropriate for smooth, non-convex optimization problems or the quasi-Bayesian approach described above.  Solving the smoothed problem can offer some computational gains relative to attempting to solve the original problem, though non-convexities remain after smoothing. The resulting estimator is first-order-equivalent to the GMM estimator for the original problem.  The estimator can, however, enjoy higher-order improved performance. \cite{kaplan2016} provide a plug-in approach to choosing the smoothing parameter $h_N$ and also demonstrate that the estimated parameters obtained from solving the smoothed problem may perform better in small samples than those from solving the unsmoothed problem or the inverse quantile regression discussed in Section \ref{Subsec: IQR}. 

\subsection{Inverse Quantile Regression}\label{Subsec: IQR}

Rather than work directly with moment condition (\ref{eq: moment}), \cite{iqr:joe} and \cite{iqr:joe2} take a different approach which they label the inverse quantile regression (IQR).  The IQR is based on the observation that (\ref{e1}) coupled with the linear quantile model (\ref{qsf: linear}) implies that the $\tau^{th}$ quantile of $Y - D'\alpha_0$ conditional on covariates $X$ and instruments $Z$ is equal to $X'\beta_0(\tau)$: 
\begin{align}\label{IQR: quantile}
Q_{Y - D'\alpha_0}(\tau|X,Z) = X'\beta_0 + Z'\gamma_0 \ \text{with} \ \gamma_0\equiv 0.
\end{align}
That is, at the true value of the coefficient vector on the endogenous variables $\alpha_0$, the conventional linear $\tau$-quantile regression of $Y-D'\alpha_0$ onto $X$ and $Z$ would yield coefficients on the instruments of exactly 0 in the population.  This observation then suggests an estimation approach based on concentrating $X$ out of the problem using conventional quantile regression, which is convex and can be solved very quickly, and then solving a lower dimensional non-convex optimization problem over only the dimension of $D$ to find $\widehat\alpha$.

Specifically, the IQR procedure works as follows.  Let $a$ denote an arbitrary hypothesized value for $\alpha_0$.  Using the hypothesized value $a$, estimate coefficients $\beta(a)$ and $\gamma(a)$ from the model $Q_{Y - D'a}(\tau|X,Z) = X'\beta(a) + Z'\gamma(a)$ by running the ordinary linear $\tau$-quantile regression of $Y - D'a$ onto $X$ and $Z$.  Let $\widehat\beta(a)$ and $\widehat\gamma(a)$ denote the resulting estimators of $\beta(a)$ and $\gamma(a)$.  Also, let $\widehat\Omega_N(a)$ denote the estimated covariance matrix of $\sqrt{N}(\widehat\gamma(a)-\gamma(a))$, and note that this covariance matrix is available in any common implementation of the ordinary quantile regression.  We can then define the IQR estimator of $\alpha_0$ as
\begin{align}\label{eq: IQR}
\widehat\alpha= \arg\min_{a\in \mathcal{A}} W_N(a),
\end{align}
where 
\begin{align}\label{eq: W_N}
W_N(a) := N\widehat\gamma(a)'\widehat\Omega_N(a)^{-1}\widehat\gamma(a).
\end{align}
Given $\widehat\alpha$, we can then estimate $\beta_0$ as $\widehat\beta(\widehat\alpha)$.  

In terms of point estimation, the main virtue of the IQR is that, by concentrating out the coefficients on exogenous variables $X$, it produces a non-convex optimization problem over only the parameters $\alpha$.  In many applications, the dimension of $D$ is small, so one can approach the non-convex optimization problem using highly robust optimization procedures that deal effectively with objectives with many local optima.  \cite{iqr:joe} recommend using a grid-search to solve (\ref{eq: IQR}) though other approaches are certainly available.  Using a grid-search is particularly appealing when coupled with weak-identification robust inference as discussed in Section \ref{Subsec: Conditional}.

\cite{iqr:joe} analyze the properties of $(\widehat\alpha(\tau)',\widehat\beta(\tau)')'$ under assumptions that guarantee strong identification.  They verify asymptotic normality of the estimator, provide a consistent estimator of the asymptotic variance, and show how instruments and observation weights can be chosen to produce an efficient estimator of the coefficients for a single quantile following \cite{chamberlain:iv}.  \cite{iqr:joe} also analyze the behavior of the process $(\widehat\alpha(\tau)',\widehat\beta(\tau)')'$ not just at a point but viewed as a function of $\tau$, providing uniform convergence results and discussing in detail applications of these convergence results to testing hypotheses about the behavior of $(\alpha_0(\tau)',\beta_0(\tau)')'$ across the index $\tau$.

\subsubsection{A Useful Interpretation of IQR as a GMM estimator.}\label{SubSub: IQRGMM}  It is useful to interpret IQR as first-order-equivalent to a particular GMM estimator, where we first profile out the coefficients on exogenous variables.

To this end, let us define 
\begin{align}\label{eq: iqr new moment}
g_\tau(V, \alpha; \beta, \delta) = \left(\tau - \mathbbm{1}( Y \leq D'\alpha + X'\beta ) \right) \Psi(\alpha, \delta(\alpha)),
\end{align}
with ``instrument"
\begin{align}\label{eq: new instrument}
\Psi(\alpha, \delta(\alpha)):= ( Z - \delta(\alpha) X). 
\end{align}
In (\ref{eq: new instrument}),
$$
\quad \delta(\alpha) = M(\alpha)J^{-1}(\alpha)
$$
where $\delta$ is a matrix parameter,
$$
{M}(\alpha) = \text{E}\left[Z  X' f_{\varepsilon}(0 \mid X,Z)\right], \quad 
{J}(\alpha) =\text{E} \left[X X' f_{\varepsilon}(0 \mid X,Z)\right],
$$ 
and $f_{\varepsilon}(0|X,Z)$ is the conditional density of $\varepsilon = Y - D'\alpha - X'\beta(\alpha)$ where $\beta(\alpha)$ is defined by
$$
\text{E}\left[(\tau - \mathbbm{1}( Y \leq D'\alpha + X'\beta(\alpha) ) X \right]=0.
$$

To proceed with estimation, for a hypothesized value $a$, we first profile out the coefficients on the exogenous variables
as in IQR,
\begin{align}\label{eq: profile}
\widehat\beta(a) = \arg\min_{b\in \mathcal{B}} \frac{1}{N}\sum_{i=1}^{N} \rho_{\tau}\left(Y_i - D_i'a - X_i'b\right).
\end{align}
We may then plug the solution of (\ref{eq: profile}) into (\ref{eq: iqr new moment}) to form 
\begin{align}\label{eq: concentrated moment}
\widehat{g}_N(a) = \frac{1}{N}\sum_{i = 1}^{N} g(V_i, 
a, \hat \beta(a),  \hat \delta(a) ),
\end{align}
where
$$
\hat \delta(a) =  \widehat{M}(a)  \widehat{J}^{-1}(a),
$$
for 
\begin{align*}
\widehat{M}(a) &= \frac{1}{N h_N}\sum_{i=1}^{N} Z_i X_i' K_{h_N}\left(Y_i - D_i'a - X_i'\widehat\beta(a)\right), \\
\widehat{J}(a) &= \frac{1}{N h_N}\sum_{i=1}^{N} X_i X_i' K_{h_N}\left(Y_i - D_i'a - X_i'\widehat\beta(a)\right),
\end{align*}
and $K_{h_N}(\cdot)$ a kernel function with bandwidth $h_N$.  
Then, we consider the GMM estimator based on the concentrated moments (\ref{eq: concentrated moment}):
$$
\hat \alpha(\tau) = \arg\min_{a \in \mathcal{A}} m_N(a),
$$
for
\begin{align}\label{eq: concentrated gmm obj}
m_N(a):= N\widehat{g}_N(a)'\widehat\Sigma(a,a)^{-1}\widehat{g}_N(a).
\end{align}
$\widehat\Sigma(a,a)$ in $m_N(a)$ is an estimator of the covariance function of the sample concentrated moment functions (\ref{eq: concentrated moment}) such as
\begin{align}\label{eq: concentrated var}
\widehat\Sigma(a_1,a_2) &= \frac{1}{N}\sum_{i=1}^{N} g\left(V_i, 
a_1, \hat \beta(a_1)\right)  g\left(V_i, 
a_2, \hat \beta(a_2)\right)'.
\end{align}
The estimator $\hat \alpha$
is first-order equivalent to the estimator $\tilde \alpha$ which employs the moment
function:
$$
g^*_\tau(V, \alpha) = (\tau - \mathbbm{1}( Y \leq D'\alpha + X'\beta_0 ) ) \Psi(\alpha_0, \delta(\alpha_0)).
$$
That is, the sample objective function for $\tilde \alpha$ uses
 \begin{align}\label{eq: concentrated moment opt}
\widehat{g}_N(\alpha) = \frac{1}{N}\sum_{i = 1}^{N} g^*_\tau(V, \alpha) 
\end{align}
where
$$
\widehat\Sigma(\alpha,\alpha) = \text{E}\left[ g_\tau^* (V_i,\alpha_0)
 g_\tau^* (V_i,\alpha_0)'\right].
$$
This equivalence holds because the moments possess the Neyman orthogonality property that we discuss later.  Moreover, by examining the first-order properties of the IQR estimator we can conclude that $\tilde \alpha$ and IQR are first-order equivalent.  

\subsection{Weak Identification Robust Inference}\label{Subsec: Conditional}

The good behavior of asymptotic approximation results for the point estimators provided in Sections \ref{Subsec: GMM}-\ref{Subsec: IQR} rely on strong identification of the model parameters as discussed in Section \ref{Subsec: Identification}. Because checking these conditions may be difficult, it is useful to have inference procedures that are robust to weak- or non-identification.

\cite{iqr:joe2} present a simple weak-identification robust inference procedure that results naturally from the IQR estimator.  The basic idea underlying this procedure is exactly the relation (\ref{IQR: quantile}) which states that the instruments $Z$ should have no explanatory power in the conventional $\tau$-quantile regression of $Y - D'\alpha_0$ on $X$ and $Z$ at the true value of the structural parameter $\alpha_0$.  Thus, a valid test of the hypothesis that $\alpha_0 = a$ for some hypothesized $a$ can be obtained by considering a test of the hypothesis that $\gamma(a) = 0$ for $\gamma(a)$ denoting the population value of the $\tau$-quantile regression coefficients defined in Section \ref{Subsec: IQR}.  Also, note that $W_N(a)$ in (\ref{eq: W_N}) is simply the standard Wald statistic for testing $\gamma(a) = 0$ and that $W_N(\alpha_0)$ converges in distribution to a $\chi^2_{\dim(Z)}$ regardless of the strength of identification of $\alpha_0$; see \cite{iqr:joe2} for details.\footnote{The same statement would also hold for the GMM objective function based on (\ref{eq: concentrated moment}) discussed in Section \ref{SubSub: IQRGMM}.}
It then follows that a valid $(1-p)\%$ confidence region for $\alpha_0$ may be constructed as the set
\begin{align}\label{CI: IQR}
\{ a \in \mathcal{A}:  W_N(a) \leq c_{1-p}\}
\end{align}
where $c_{1-p}$ is such that $P\left[\chi^2_{\dim(Z)} > c_{1-p}\right] = p$, and the set may be approximated numerically by considering $a$'s in the grid $\{a_j, j =1,...,J\}$.  Thus, a natural byproduct of solving (\ref{eq: IQR}) through a grid search is a confidence set for the structural parameter $\alpha_0$ that is valid regardless of the strength of identification of the parameter.  We note that this procedure could also be adapted to be used with the orthogonal scores defined in Section \ref{Subsec: Neyman} to provide weak-identification robust inference in settings with high-dimensional $X$ or other settings where robustness to estimation of the nuisance parameter $\beta_0$ is a major concern.

The approach of \cite{iqr:joe2} outlined above is in the spirit of the weak identification robust procedure of \cite{anderson:rubin}.  The procedure is relatively simple to implement, but suffers from the same well-known lack of power as other Anderson-Rubin-type statistics in overidentified models under strong identification.  To overcome this potential inefficiency, \cite{JunWeakIdIVQR} proposes a different statistic analogous to the proposal of \cite{kleib:weakgmm} which is locally efficient under strong identification but may suffer from substantial declines in power against alternatives that are distant from the true parameter value.  In the following, we discuss the related approach of \cite{AM:functionalnuisance} which extends the conditional likelihood ratio approach of \cite{moreira:weakivlr} to general nonlinear settings.  This approach retains efficiency under strong identification but also maintains good power against distant alternatives.

The \cite{AM:functionalnuisance} approach employs a quasi-likelihood ratio (QLR) statistic as
\begin{align}\label{eq: QLR}
QLR_N(a) = m_N(a) - \inf_{a \in \mathcal{A}} m_N(a)
\end{align}
where $m_N(a)$ is the GMM objective function (\ref{eq: concentrated gmm obj}). 

Under weak identification, the distribution of $QLR_N(a)$ is non-standard and depends on a nuisance function that is not consistently estimable.  \cite{AM:functionalnuisance} provide a sufficient statistic (in LeCam's Gaussian limit experiment)
$$
S(a) = \sqrt{N}\left(\widehat{g}_N(a) - \widehat\Sigma(a,\alpha_0)\widehat\Sigma(\alpha_0,\alpha_0)^{-1}\widehat{g}_N(\alpha_0)\right)
$$ 
for this functional nuisance parameter, where $g_N(a)$ and $\widehat\Sigma(a_1,a_2)$ are defined in (\ref{eq: concentrated moment}) and (\ref{eq: concentrated var}).  \cite{AM:functionalnuisance} also outline a procedure to simulate the distribution of $QLR(a)$ conditional on $S(a)$ that proceeds as follows. First, draw $\zeta^*_b \sim N\left(0,\widehat\Sigma(\alpha_0,\alpha_0)\right)$ for $b = 1,...,B$ for a large number $B$.  For each $\zeta^*_b$, the QLR statistic for that draw is then calculated as 
$$
QLR_{N,b}^*(a) = m_{N,b}^*(a) - \inf_{a \in \mathcal{A}} m_{N,b}^*(a)
$$
where 
$$
m_{N,b}^*(a) = N\widehat{g}_{N,b}^*(a)'\widehat\Sigma(a,a)^{-1}\widehat{g}_{N,b}^*(a)
$$
for
$$
\widehat{g}_{N,b}^*(a) = S(a) + \widehat\Sigma(a,\alpha_0)\widehat\Sigma(\alpha_0,\alpha_0)^{-1}\zeta^*_b.
$$
The simulated distribution then provides an appropriate critical value, $c_{1-p}(S(a))$, for performing a valid $p$-level test of the null hypothesis that $\alpha_0= a$ by rejecting when $QLR_N(a) > c_{1-p}(S(a))$.  It then follows that a valid $(1-p)\%$ confidence region for $\alpha_0$ is given by 
$$
\{ a \in \mathcal{A}:
QLR_N(a) \leq c_{1-p}(S(a))\}.
$$

\subsection{Finite Sample Inference}

The inference procedures reviewed in the previous sections all rely on asymptotic approximations. \cite{fsqr} provide a finite sample inference approach which can also be used if the validity of the assumptions necessary to justify these approximations is questionable and is valid in setups with weak or set identification.

Their approach makes use of the fact under the assumptions of the IVQR model, the event $\left\{Y\le q\left(\tau,D,X\right) \right\}$ conditional on $(Z,X)$ is distributed exactly as a Bernoulli$(\tau)$ random variable regardless of the sample size. This random variable depends only on $\tau$, which is known, and so is pivotal in finite samples. For the GMM objective function $m_N\left(\theta_0\right)$ defined in (\ref{eq: GMM obj}), this implies that $m_N\left(\theta_0\right)\overset{d}{=}\widetilde{m}_N$ conditional on $\{X_i,Z_i\}_{i=1}^{N}$, where
$$
\widetilde{m}_N:=\left(\frac{1}{\sqrt{N}} \sum_{i=1}^N\left( \tau-B_i\right)\cdot \Psi_i\right)'\Omega_N \left( \frac{1}{\sqrt{N}}\sum_{i=1}^N\left( \tau-B_i\right)\cdot \Psi_i\right)
$$
and $\{B_i\}_{i=1}^{N}$ are i.i.d. Bernoulli random variables that are independent of $\{X_i,Z_i\}_{i=1}^{N}$ and have $\text{E}\left[B_i\right]=\tau$. This result provides the finite sample distribution of the GMM function  $m_N\left(\theta\right)$ at $\theta=\theta_0$, which does not depend on any unknown parameters. Given the finite sample distribution of $m_N\left(\theta_0\right) $, a $p$-level test of the null hypothesis that $\theta=\theta_0$ is given by the rule that rejects the null if $m_N\left(\theta\right)>c_{1-p}$, where the critical value $c_{1-p}$ is the $(1-p)^{\text{th}}$ quantile of $\widetilde{m}_N(\tau)$. It then follows that a valid $(1-p)\%$ joint confidence set for $\theta$ is given by
$$
\{ \theta \in \Theta:
m_N\left(\theta\right) \leq c_{1-p}\}.
$$
We note that inference is simultaneous on all components of $\theta$ and that for joint inference the approach is not conservative. Inference about subcomponents of $\theta$ such as $\alpha$ may be made by projections and may be conservative.

The chief difficulty with the finite sample approach is computational. Implementing the approach requires inversion of the function $m_N\left(\theta\right) $, which may be quite difficult if the number of parameters is large. To alleviate this problem, \cite{fsqr} develop suitable MCMC algorithms.

\section{Advanced Inference with High-Dimensional X}

\subsection{Neyman-Orthogonal Scores}\label{Subsec: Neyman}

Here we deal with the case where we have high-dimensional covariates. Such cases are common in current high-dimensional data sets where one may see very many potential control variables.  High-dimensional covariates also arises in semiparametric problems; for example, we may be interested in a partially linear structural quantile model
\begin{align*}
q(\tau,d,w) = \alpha_0(\tau)d + g(\tau,w)
\end{align*}
where $W$ is a low-dimensional set of variables and we approximate $g(\tau,w) \approx x'\beta_0(\tau)$ using a collection of approximating functions $x = h(w)$.  In settings with high-dimensional $X$, estimation of $\beta_0(\tau)$ may contaminate estimation of the parameters of interest, $\alpha_0(\tau)$, leading to a breakdown of estimation and inference based directly on (\ref{eq: moment}).  The potential for contamination is especially acute in high-dimensional settings where some form of regularization will be used to make informative estimation feasible but may arise more generally.

Due to the potentially poor finite sample performance of estimators based directly on (\ref{eq: moment}), one might prefer to base estimation and inference on ``orthogonal'' moment conditions that are relatively insensitive to estimation of the nuisance parameters $\beta_0$.  Specifically, we may prefer to base estimation and inference for $\alpha_0$ on moment functions 
$$
g(V, \alpha; \eta), \text{ where }  V = (Y,D,Z,X)
$$
and $\eta$ denotes nuisance parameters with true values $\eta_0$ that include $\beta_0$ as a sub-component, that identify $\alpha_0$ via
\begin{align}\label{eq: new moment}
\text{E}[g(V,\alpha_0;\eta_0)] = 0
\end{align}
and obey the Neyman orthogonality condition:
\begin{align}\label{eq: orthogonal}
\partial_{\eta} \text{E}[g(V, \alpha_0; \eta)] \Big |_{\eta= \eta_0} = 0
\end{align}
where $\partial_\eta$ denotes a functional derivative operator.  (\ref{eq: orthogonal}) is the key orthogonality condition that ensures that the moment conditions defining $\alpha_0$ are locally insensitive to perturbations in the nuisance parameters.  This property results in the first-order properties of estimation and inference of $\alpha_0$ based on sample analogs to (\ref{eq: new moment}) being insensitive to estimation of nuisance functions as long as sufficiently high-quality estimators of the nuisance functions are available.  The idea of using orthogonal estimating equations goes back at least to \cite{Neyman59} and \cite{Neyman1979} where they were used in construction of Neyman's celebrated $C(\alpha)$-statistic.  The use of moment conditions satisfying the orthogonality condition (\ref{eq: orthogonal}) is crucial for establishing good properties of semi-parametric estimators in modern, high-dimensional estimation settings when regularized estimation or other machine learning tools are used in estimation of nuisance functions; see, e.g. \cite{BCFH:Policy}, \cite{CHS:AnnRev}, and \cite{DoubleML}.  

The  orthogonal moment functions for the IVQR setting are given by
$$
g_\tau(V, \alpha, \eta) = (\tau - \mathbbm{1}( Y \leq D'\alpha + X'\beta ) ) \Psi(\alpha, \delta(\alpha)),
$$
where  $\Psi(\alpha, \delta(\alpha))$ and $\delta(\alpha)$ are defined in Section \ref{SubSub: IQRGMM}.
The nuisance parameter and its true value are then given by
$$
\eta : = (\beta,  \delta(\alpha)), ~~\text{and}~~ \eta_0:=  (\beta_0,  \delta(\alpha_0)).
$$

Observe that the Neyman orthogonality condition holds for these moment conditions because, under appropriate smoothness conditions,
$$
\partial_{\beta} \text{E}[g(V, \alpha_0; \eta] \Big |_{\eta= \eta_0} = 
 M(\alpha_0) -  M(\alpha_0)J^{-1}(\alpha_0) J(\alpha_0) = 0,
$$
$$
\partial_{\delta} \text{E}[g(V, \alpha_0; \eta] \Big |_{\eta= \eta_0}
= \text{E} \left[ (\tau - \mathbbm{1}( Y \leq D'\alpha_0 + X'\beta_0 ) ) X\right] = 0.  
$$

\subsection{Estimation and Inference Using Orthogonal Scores}

We start similarly to the IQR estimator by first profiling out the coefficients on exogenous variables using an $\ell_1$-penalized quantile regression estimator to define
\begin{align}\label{eq: profile2}
\widehat\beta(a) = \arg\min_{b\in \mathcal{B}} \frac{1}{n}\sum_{i=1}^{N} \rho_{\tau}(Y_i - D_i'a - X_i'b)
+ \lambda \sum_{j=1}^{\dim(b)}  \psi_j | b_j |.
\end{align}
for a hypothesized value $a$. We then estimate 
\begin{align*}
\widehat{M}(a) &= \frac{1}{N h_N}\sum_{i=1}^{N} Z_i X_i' K_{h_N}\left(Y_i - D_i'a - X_i'\widehat\beta(a)\right), \\
\widehat{J}(a) &= \frac{1}{N h_N}\sum_{i=1}^{N} X_i X_i' K_{h_N}\left(Y_i - D_i'a - X_i'\widehat\beta(a)\right),
\end{align*}
for $K_{h_N}(\cdot)$ a kernel function with bandwidth $h_N$ as before.   Since 
$\widehat{J}(a)$ is high-dimensional and is not invertible, we may estimate 
row-components $\delta_j(a)$ of matrix $\delta(a)$ by solving the $\ell_1$-regularized problem
$$
\hat \delta_j(a) = \arg \min_{\delta} \frac{1}{2} \delta' \hat J(a) \delta - \hat M_j(a) \delta + \vartheta \|\delta\|_1,
$$
where $\hat M_j(a)$ is the $j$-th row of $\hat M(a)$, interpreted as a row vector itself, and $\vartheta$ is a penalty level.  The solution $\hat \delta_j(a)$ obeys the Karush-Kuhn-Tucker condition
\begin{equation}\label{KKT}
\|\hat \delta_j(a)' \hat J(a)  - \hat M_j(a)\|_{\infty} \leq \vartheta,  \forall j,
\end{equation}
so we may think of $\hat \delta_j(a)$ as a regularized estimator of $M_j(a)J^{-1}(a)$.

Alternatively we can the regularized estimator via Dantzig form of Lasso
by minimizing a norm of $\hat \delta(a)$ subject to the above constraints (\ref{KKT}).

We may then plug in the solution of (\ref{eq: profile2}) to form a concentrated sample moment function analogous to (\ref{eq: sample moment}) as
\begin{align}\label{eq: concentrated moment orth}
\widehat{g}_N(a) = \frac{1}{N}\sum_{i = 1}^{N} \left(\tau - \mathbbm{1}\left(Y_i - D_i'a - X_i'\widehat\beta(a) \leq 0\right)\right) \Psi ( a, \hat \delta (a) ).
\end{align}
These concentrated moments can be used to set-up the continuously-updated GMM estimator:
$$
\hat \alpha = \arg\min_{a \in \mathcal{A}} N\widehat{g}_N(a)'\widehat\Sigma(a,a)^{-1}\widehat{g}_N(a),
$$
where again $\widehat\Sigma(a,a)$ is an estimator of the covariance function of the sample concentrated moment functions (\ref{eq: concentrated moment orth}).  The estimator $\widehat\alpha$ would then follow standard properties of the infeasible GMM estimator that replaced the estimators $\hat \beta(a)$ and $\hat \delta (a)$ with their true values $\beta_0$ and $\delta(\alpha_0)$ as long as instruments are low dimensional and identification is strong.  If the set of instruments was also high-dimensional, further regularization would be called for to make reliable estimation and inference feasible.

We can also directly use the concentrated moments to set-up standard Anderson-Rubin-type inference for $\alpha_0$ under weak or partial identification as in Section \ref{Subsec: Conditional}.  Similarly, we could base inference from more refined approaches, such as \cite{AM:functionalnuisance}, on the concentrated moments.  Indeed, we can use these concentrated moments  to form a quasi-likelihood ratio (QLR) statistic as
\begin{align}\label{eq: QLR2}
QLR_N(a) = N\widehat{g}_N(a)'\widehat\Sigma(a,a)^{-1}\widehat{g}_N(a) - \inf_{a \in \mathcal{A}} N\widehat{g}_N(a)'\widehat\Sigma(a,a)^{-1}\widehat{g}_N(a).
\end{align}
Because of the orthogonality property, estimation of the nuisance parameters does not affect the first-order behavior of the empirical moments, so inference based on (\ref{eq: QLR2}) falls back exactly in the setting of \cite{AM:functionalnuisance}.  One could then employ their approach
to compute the critical values for $QLR_N(a)$ conditional on a sufficient statistic, $c_{1-p}(QLR_N(a))$.
It then follows that a valid $(1-p)\%$ confidence region for $\alpha_0$ may be constructed by considering $a$'s in the grid $\{a_j, j =1,...,J\}$ exactly as in approximating (\ref{CI: IQR}).

\section{Conclusion}\label{Sec:Conclusion}

In this chapter, we have reviewed the structural IVQR model developed in \cite{iqr:ema} which can be used to estimate causal quantile effects in the presence of endogeneity.  The model makes use of instrumental variables that satisfy conventional independence and relevance conditions from the nonlinear instrumental variables literature.  Specifically, instruments are assumed to be independent of unobservables associated to potential outcomes but related to endogenous right-hand-side variables in the model.  The presence of instruments alone is insufficient to identify QTE, and the IVQR models imposes an additional condition on structural unobservables, termed rank similarity, that restricts the distribution of unobservables in potential outcomes across different potential states of the endogenous variables.  Under these conditions, an IV-style moment condition can be derived which then provides a basis for identification and estimation of QTE.  We provided two concrete examples of economic models that fall within the IVQR framework. 

We then reviewed leading approaches to estimating model parameters and performing inference for QTE within the IVQR model based on the moment conditions implied by the model.  Estimation and inference is complicated by the non-smooth and non-convex nature of the IVQR moment conditions.  We discuss estimation and inference approaches that attempt to alleviate this issue.  We also review approaches to inference which remain valid under weak or even non-identification.

There are, of course, many open areas for research in quantile models with endogeneity. As discussed in Section \ref{SubSec: Other Approaches}, \cite{abadie} and \cite{imbens:newey} offer alternative approaches to identifying QTE by imposing alternate sets of assumptions to those used in the IVQR model. These approaches and the IVQR model are non-nested and further understanding their connections may be interesting.  \cite{wuthrich:ivqr} provides a contribution in this direction by showing the connection between the estimands of both models within the structure of the \cite{abadie} framework. It would also be interesting to analyze the properties of the IVQR estimands when some of the underlying assumptions are violated. Towards this end, \cite{wuthrich:ivqr} provides a characterization of QTE estimands based on the IVQR model with binary treatments in the absence of rank similarity.
Another topic that may deserve further consideration is the systematic analysis of estimation and inference based on the orthogonal moment equations sketched in Section \ref{Subsec: Neyman}, especially in high-dimensional settings.  We also note that the IVQR model may be useful for uncovering structural objects even if quantile effects are not the chief objects of interest; see, for example, \cite{BerryHaileDiscreteChoiceId}.  It may be interesting to further explore application of the IVQR model and related estimation methods in structural economic applications.  Finally, a potentially interesting but more unexplored area may be to think about quantile-like quantities for multivariate outcomes with endogenous covariates.

\bibliographystyle{elsarticle-harv}
\bibliography{IVQRchapterBib}

\printindex

\end{document}